\documentclass{article}
\usepackage{spconf,amsmath,graphicx,hyperref}
\usepackage{url}
\usepackage{booktabs,adjustbox,makecell,multirow}
\usepackage{framed}

\title{The USTC-NERCSLIP Systems for the CHiME-9 MCoRec Challenge}
 \name{\parbox{\linewidth}{\centering
    Ya Jiang$^{1}$,
	Ruoyu Wang$^{1}$,
	Jingxuan Zhang$^{2}$,
	Jun Du$^{1*}$,
	Yi Han$^{1}$,
	Zihao Quan$^{1}$,
    Hang Chen$^{1}$,
    Yeran Yang$^{2}$, 
    Kongzhi Zheng$^{2}$,
    Zhuo Chen$^{3}$,
    Yanhui Tu$^{4}$,
    Shutong Niu$^{5}$,
    Changfeng Xi$^{5}$,
    Mengzhi Wang$^{5}$, 
    Zhongbin Wu$^{5}$,
    Jieru Chen$^{5}$,
    Henghui Zhi$^{5}$, 
    Weiyi Shi$^{5}$, 
    Shuhang Wu$^{5}$,
    Genshun Wan$^{5}$,
    Jia Pan$^{5}$,
    Jianqing Gao$^{5}$
	\thanks{$^{*}$Corresponding author.}}}
\address{$^1$University of Science and Technology of China, Hefei, China \\
	$^2$Shaanxi Normal University, Xi'an, China $^3$Lomonosov Moscow State University, Moscow, Russia  \\
        $^4$Anhui University, Hefei, China  $^5$iFLYTEK Research, iFLYTEK Co., Ltd., Hefei, China }
\begin{document}
%
\maketitle
\begin{abstract}
This report details our submission to the CHiME-9 MCoRec Challenge on recognizing and clustering multiple concurrent natural conversations within indoor social settings. Unlike conventional meetings centered on a single shared topic, this scenario contains multiple parallel dialogues--up to eight speakers across up to four simultaneous conversations--with a speech overlap rate exceeding $90\%$. To tackle this, we propose a multimodal cascaded system that leverages per-speaker visual streams extracted from synchronized 360$^\circ$ video together with single-channel audio. Our system improves three components of the pipeline by leveraging enhanced audio-visual pretrained models: Active Speaker Detection (ASD), Audio-Visual Target Speech Extraction (AVTSE), and Audio-Visual Speech Recognition (AVSR). The AVSR module further incorporates Whisper and LLM techniques to boost transcription accuracy. Our best single cascaded system achieves a Speaker Word Error Rate (WER) of $32.44\%$ on the development set. By further applying ROVER to fuse outputs from diverse front-end and back-end variants, we reduce Speaker WER to $31.40\%$. Notably, our LLM-based zero-shot conversational clustering achieves a speaker clustering F1 score of 1.0, yielding a final Joint ASR-Clustering Error Rate (JACER) of $15.70\%$.
\end{abstract}
%

\section{System Description}
\label{sec:system}
Our system independently processes the official session-level cropped central tracks in the CHiME-9 MCoRec Challenge\cite{nguyen2025cocktailtask, nguyen2025cocktailbenchmark}. First, an active speaker detection (ASD) system is used to estimate frame-level speaker activity probabilities for each track video, and segmentations are generated based on the given onset and offset thresholds.  Next, audio-visual target speech extraction (AVTSE) is performed on each segmented region and the extracted speech is then fed into the audio-visual speech recognition (AVSR) system. Finally, the conversation clustering system groups speakers into their corresponding conversations.

\subsection{ASD}
\label{ssec:ASD}
Our ASD module leverages a pre-trained audio-visual encoder (base full-face encoder in Section\ref{ssec:AVSR}) to exploit robust cross-modal representations. 
The architecture integrates a primary binary classifier for target detection and an auxiliary visual classifier to reinforce modality-specific discrimination, both optimized via Binary Cross-Entropy (BCE) loss. 
To balance task adaptation with the preservation of pre-trained knowledge, we employ a two-stage transfer learning strategy: initially fine-tuning the classifiers with a frozen encoder, followed by end-to-end optimization of the entire network.

We improved the robustness of the model through three key mechanisms to address the complex acoustic scenarios of CHiME-9: (1) \textbf{Data Specification}: We incorporate a diverse set of auxiliary datasets, including the CHiME-9 official dataset, the AVA dataset~\cite{AVA}, the MSDWILD dataset~\cite{MSDWild}, and the M3SD dataset~\cite{M3SD}. We adopt a unified preprocessing pipeline that trims invalid segments, applies geometric face standardization and re-tracking to ensure spatial consistency, and uses annotations to extract speaker-wise audio and face tracks. The extracted tracks are further sliced into short segments for training, resulting in approximately 200 hours of audio-visual ASD data; (2) \textbf{Online Augmentation}: We simulated multi-speaker environments (2--4 participants) via aggressive noise injection; and (3) \textbf{Temporal Smoothing}: We developed a post-processing pipeline that applies softmax normalization and merges fragmented predictions to ensure physically consistent speaker activity timelines.

\subsection{AVTSE}
\label{ssec:AVTSE}  
The CHiME-9 MCoRec Challenge features extremely severe acoustic interference, with speech being persistently suppressed by competing dialogues. We develop four AVTSE systems (not present in the official baseline) that exploit distinct self-supervised pretrained models and multimodal architectures to disentangle overlapping speech streams, and ensemble the epoch-level mask averaging per system.

\textbf{Data Simulation:} We constructed synthetic training sets using  public audio-visual corpora---LRS3~\cite{afouras2018lrs3}, VoxCeleb2~\cite{Chung18b}, and AVSpeech~\cite{10.1145/3197517.3201357}---augmented with the DNS-Noise~\cite{dubey2023icassp} and preprocessed as detailed in Section\ref{ssec:AVSR}. We performed intra-dataset simulation, reserving 10\% of speakers as disjoint interference sources. Interference profiles comprised 45\% single, 45\% dual, and 10\% noise-only sources, mixed at SNRs uniformly sampled from $[-10, 10]$ dB. The raw synchronized audio-visual segment serve as the clean target speech and its corresponding visual cue. Although we initially modeled CHiME-9's dense overlap (up to 7 speakers), empirical evaluations indicated that training on such high-complexity data degraded generalization. This yielded 2–3 speaker mixed datasets of 335 hours (LRS3), 1150 hours (VoxCeleb2), and 1421 hours (AVSpeech), with the latter incorporated exclusively for \textbf{System IV}.

\textbf{System Configurations:} \textbf{System I} establishes the AVTSE baseline, employing BRAVEn~\cite{haliassos2024braven} encoders (a self-supervised framework fine-tuned on LRS3 for ASR/VSR tasks) to extract semantic-phonetic representations from raw audio mixture and target lip sequences. These features, concatenated with Log Power Spectrum (LPS) processed by a squeezed TCN (S-TCN) and residual LSTM, are fed into a Conformer network for temporal modeling. Then, a S-TCN decoder predicts the Ideal Ratio Mask (IRM) optimized via MSE loss. \textbf{System II} augments the baseline architecture with a large full-face encoder pre-trained on large-scale paired audio-visual data in Section~\ref{ssec:AVSR}, forming a tri-modal framework that captures explicit speaker identity cues beyond phonetic content from BRAVEn. \textbf{System III} introduces joint TSE-ASR optimization: reconstructed speech is encoded by \textbf{AVSR S3} encoder, and a cosine similarity loss between reconstructed and clean features serves as a regularization term, implicitly enhancing the intelligibility of the reconstructed speech. \textbf{System IV} adopts an alternative dual-tower\cite{chen2021correlating} with parallel ResNet-18 backbones for audio and visual streams, fused via a Multi-Scale TCN, pre-trained to correlate phonemes and visemes. The fused multimodal embedding is integrated with LPS features following \textbf{System I}' paradigm, offering a feature space complementary to the BRAVEn-based approaches.

\vspace{-0.3cm}
\subsection{AVSR}
\label{ssec:AVSR}
\begin{figure}
\centering
\includegraphics[width=\linewidth]{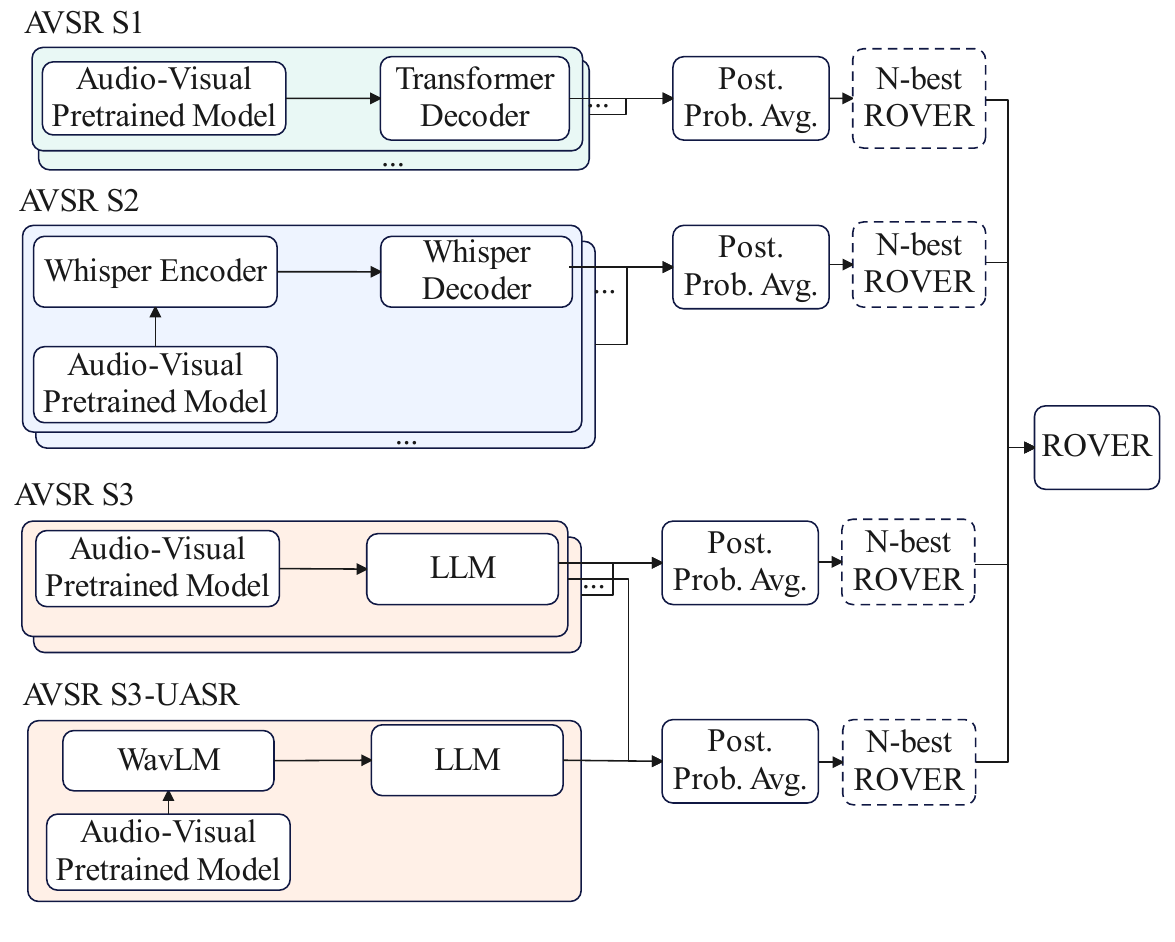}
\vspace{-0.8cm}
\caption{Overview of our audio-visual speech recognition methods. N-best ROVER is only used in our submission 3.}
\label{fig:avsr}
\vspace{-0.3cm}
\end{figure}
To ensure robust transcription in complex acoustic environments, we developed a diverse ensemble of different AVSR frameworks centered around a novel self-supervised pre-training strategy, including fine-tuning with a randomly initialized decoder (\textbf{AVSR S1}), fusion with Whisper~\cite{radford2023robust} (\textbf{AVSR S2}), and integration with a large language model (LLM)-based decoder (\textbf{AVSR S3} and \textbf{AVSR S3-UASR}). The final transcription is generated by hierarchically combining the outputs of three distinct frameworks via posterior probability averaging and ROVER~\cite{659110} method.

\textbf{Data Preparation:} In addition to the CHiME dataset, we use five publicly available audio-visual datasets for AVSR training: the pretraining (195 h) and training (28 h) subsets of LRS2~\cite{afouras2018deep}; the pretraining (408 h) and train–validation (30 h) subsets of LRS3~\cite{afouras2018lrs3}; the English portion of VoxCeleb2~\cite{Chung18b} following Shi et al., resulting in 1,326 hours (with text labels transcribed by Whisper-large-v2); AVSpeech~\cite{10.1145/3197517.3201357}, which is segmented and preprocessed following the protocols of LRS2 and LRS3, and then language-identified and transcribed using Whisper-large-v2, where only the English portion is retained, yielding 1,436 hours for training; and the talking-face subset of AVYT~\cite{nguyen25b_interspeech} (1,449 h).
After preprocessing, excluding AVYT, we obtain a total of 3,530 hours of audio-visual data containing both full-face and lip ROI videos. Incorporating AVYT further adds 1,436 hours of lip-only video data, resulting in 4,966 hours of audio-visual training data in total.

\textbf{Audio-Visual Pretraining:}
Our backbone model employs a masked-prediction objective inspired by DistillAV~\cite{zhang2025audio}, with three critical architectural enhancements. First, we integrate a ConvNeXt~\cite{liu2022convnet} frontend with Enhanced Conformer blocks~\cite{gulati20_interspeech}. Second, deviating from mel-spectrogram targets, we adopt a BEST-RQ~\cite{chiu2022self} style loss: the model predicts discrete codes (vocabulary size 8,192) derived by quantizing the hidden representations of a frozen Whisper model. Third, to capture broader facial cues beyond the lip region based on our previous studies~\cite{zhang2022lip, zhang2025target}, we utilize dual visual inputs---full-face and lip-ROI---processed by separate frontends and concatenated along the feature dimension.

\textbf{System Configurations:} \textbf{AVSR S1} attaches a randomly initialized Transformer decoder to the pre-trained encoder. The entire network is fine-tuned using a hybrid CTC/Attention loss. To mitigate overfitting, we apply adaptive temporal masking for video and SpecAugment for audio. 
\textbf{AVSR S2} modelizes the Whisper~\cite{radford2023robust} backbone. We utilize our pre-trained model (video branch only) to extract visual representation, which is injected into the Whisper encoder via Flamingo-style~\cite{NEURIPS2022_960a172b, rouditchenko24_interspeech} cross-attention layers inserted into every odd-numbered Transformer block, allowing the frozen Whisper backbone to explicitly attend to visual cues. \textbf{AVSR S1} and \textbf{AVSR S2} are first fine-tuned on internal datasets before domain adaptation to CHiME-9.
\textbf{AVSR S3} projects the output of the fine-tuned S1 encoder into the token embedding space of Qwen-2.5~\cite{qwen2025qwen25technicalreport}. We employ a staged training protocol: initially training the adaptor, then jointly optimizing with Low-Rank Adaptation (LoRA)~\cite{hu2022lora} parameters injected into the LLM. Additionally, \textbf{AVSR S3-UASR} implements our previous UASR-LLM framework~\cite{zhang2025adapting}, where visual features are injected into a WavLM encoder via Flamingo layers before being projected to the LLM. The model is trained using a two-stage strategy consisting of visual injection pretraining and AVSR fine-tuning.

\textbf{Model Ensembling:}
We employ a two-level fusion strategy. First, models sharing the same tokenizer (within-system) are combined via \textit{posterior probability averaging} to stabilize predictions. Second, the final submission is derived using ROVER to aggregate the text-level outputs across the heterogeneous architectures \textbf{(AVSR S1--S3)} and their N-best beam search hypotheses.

\begin{figure}
\centering
\includegraphics[width=\linewidth]{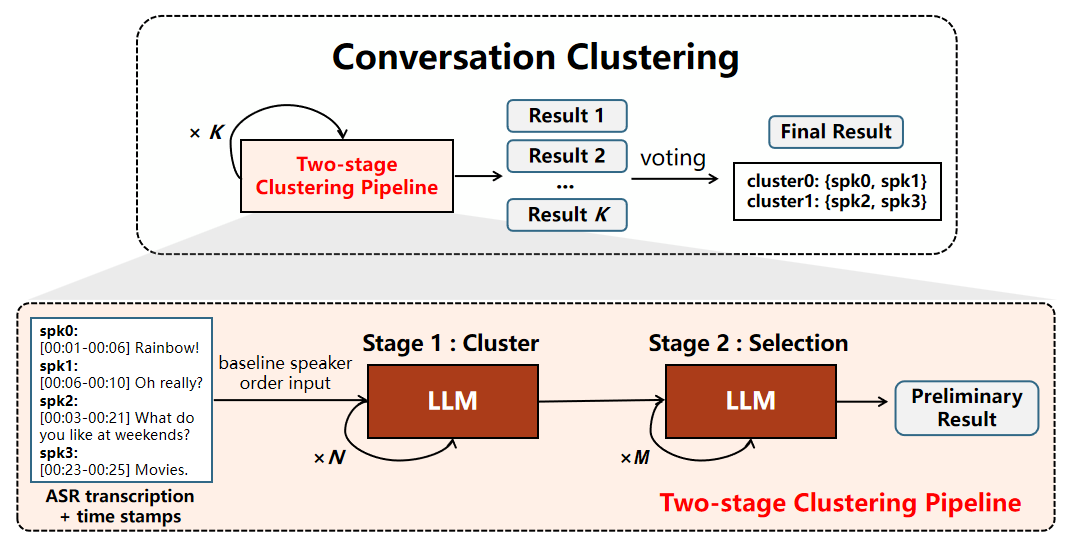}
\caption{Overview of our conversation clustering method.}
\label{fig:sc}
\end{figure}

\subsection{Conversation Clustering}
\label{ssec:cc}
Semantic cues---such as topic coherence and question-answer dependencies---are essential for grouping speakers into distinct conversation clusters. To exploit these semantic features, we leverage large language models (LLM), including Qwen 2.5 (70B) \footnote{https://huggingface.co/Qwen} and DeepSeek R1 (671B) \footnote{https://huggingface.co/deepseek-ai}, which are highly proficient in context understanding, to infer speaker relationships directly from ASR transcripts. We propose a robust two-stage ensemble pipeline as shown in Fig.~\ref{fig:sc}. At the initial \textit{Cluster Stage}, LLM is prompted independently for $N$ times to produces diverse candidate speaker-to-conversation assignments, with ASR transcriptions, timestamps, and a baseline speaker order as inputs. 
In the \textit{Selection Stage}, the model identifies the optimal clustering result from the candidates. To filter out random errors, we repeat this selection $M$ times and choose the most frequent outcome. Finally, the entire two-stage pipeline is executed $K$ times, followed by a voting fusion. 
The final submission aggregates predictions from the baseline, Qwen, and DeepSeek systems. Due to the substantial computational and time cost of the 671B parameter model, DeepSeek-R1 inference was constrained to limited iterations ($M=5, K=5$). The specific prompt designs are detailed as follows: (1) \textbf{Stage 1 Prompt -} Please perform conversation clustering based on the transcription results. You may make judgments based on factors such as conversation content, shared topics, dialogue structure, and temporal overlap between utterances; (2) \textbf{Stage2 Prompt -} Please select the most plausible result from the candidate clustering outputs. Your decision may be based on factors such as conversation content, shared topics, dialogue structure, and temporal overlap between utterances. From these candidate results, select the most likely one.

\begin{table}[t]
\centering
\vspace{-0.2cm}
\caption{WER(\%) results of submissions on $\mathcal{D}_{\mathrm{dev}}$ and $\mathcal{D}'_{\mathrm{dev}}$.}
\begin{tabular}{c|ccc}
\hline
\textbf{Set} & \textbf{Submission 1} & \textbf{Submission 2} & \textbf{Submission 3} \\
\hline
$\mathcal{D}_{\mathrm{dev}}$  & 32.03 & 31.43 & \textbf{31.39} \\
$\mathcal{D}'_{\mathrm{dev}}$ & 29.98 & 29.36 & \textbf{29.35} \\
\hline
\end{tabular}
\label{tab:dev_results}
\vspace{-0.3cm}
\end{table}

\vspace{-0.3cm}
\section{Results}
We denote the refined development set as \(\mathcal{D}'_{\mathrm{dev}}\), derived from the original set \(\mathcal{D}_{\mathrm{dev}}\) by excluding three samples: Session 53 (Speaker 2), Session 54 (Speaker 4), and Session 55 (Speaker 3). These instances were omitted due to identified misalignment between the annotations and the corresponding video.

\vspace{-0.4cm}
\subsection{Overall Results}

All AVSR systems mentioned have already been internally fused via posterior probability averaging, as described in Section~\ref{ssec:AVSR} (Model Ensembling). Detailed posterior probability averaging results are provided in Table~\ref{tab:avsr}. The construction details of our three submissions are described as follows.
\textbf{Submission 1}:
The output hypotheses from \textbf{AVSR S1, S2, S3} and \textbf{S3-UASR} based on the original audio inputs are fused using ROVER.
\textbf{Submission 2}:
For both the original audio and four speech separation systems, we generate 20 decoding results using \textbf{AVSR S1, S2, S3} and \textbf{S3-UASR}. A heuristic ROVER-based search strategy is applied to identify the optimal system combination. The selected AVTSE–AVSR combinations include the following nine systems:
\textbf{System III - S3-UASR}, 
\textbf{System IV - S3-UASR}, 
\textbf{System II - S3}, 
\textbf{System IV - S3}, 
\textbf{Original - S3}, 
\textbf{System III - S2}, 
\textbf{System IV - S2}, 
\textbf{System II - S1}, 
and \textbf{Original - S2}.
\textbf{Submission 3}:
For the nine systems described in Submission 2, we first perform single-system ROVER fusion with N-best=30 for each system individually. The resulting nine fused hypotheses are then further combined using ROVER to obtain the final output.
The detailed evaluation results are presented in Table~\ref{tab:dev_results}.

\begin{table}[t]
\centering
\vspace{-0.1cm}
\caption{Recall and Precision results of the ASD model on $\mathcal{D}_{\mathrm{dev}}$ and the corresponding WER(\%) on $\mathcal{D}'_{\mathrm{dev}}$.}
\begin{tabular}{c c c c}
\hline
\textbf{System} & \textbf{Recall (\%)}  & \textbf{Precision (\%)}  & \textbf{WER(\%)} \\
\hline
baseline  &  75.51  & 94.95 & 33.62 \\
\hline
Ours   &  \textbf{82.74}  & \textbf{95.92} & \textbf{31.23} \\
\hline
\end{tabular}
\label{tab:ASD}
\end{table}

\begin{table}[t]
\centering
\vspace{-0.2cm}
\caption{WER(\%) results across different AVTSE systems implemented upon our ASD and \textbf{AVSR S3} system on $\mathcal{D}'_{\mathrm{dev}}$.}
\begin{tabular}{c|ccccc}
\hline
\textbf{System} & \textbf{I} & \textbf{II} & \textbf{III} & \textbf{IV} & \textbf{Original} \\
\hline
WER & 30.90 & 30.82 & 30.89 & 31.04 & 31.23 \\
\hline
\end{tabular}
\label{tab:AVTSE}
\end{table}

\subsection{Detail Results}
For ASD, the experimental results are reported in Table~\ref{tab:ASD}. Compared to the baseline model light-ASD~\cite{lightASD}, our model shows significant improvements in both recall and precision, with absolute increases of 7.23\% and 0.97\%, respectively.

For AVTSE, we compared the four AVTSE systems on our own ASD and \textbf{AVSR S3} model as shown in Table~\ref{tab:AVTSE}. Experimental results indicate that audio extracted through different AVTSE systems all yielded varying degrees of improvement compared to the original mixture input, whereas the popular extraction schemes---such as AV-Speformer, USEV and Seanet---all resulted in performance degradation under such challenging conditions.

For AVSR, we evaluated two pre-training capacities—\textbf{Base} (12 layers, 110M params) and \textbf{Large} (18 layers, 330M params)—across three visual configurations: \textbf{Face} (full-face), \textbf{Lip} (lip-ROI), and \textbf{Face\&Lip}. Face-dependent models utilized a 3530-hour subset (excluding AVYT), while lip-based models leveraged the full 4966-hour corpus. Evaluation was conducted on \(\mathcal{D}'_{\mathrm{dev}}\) using official ASD segmentation (concatenated to 8--16s) and raw far-field audio, without N-best ROVER.
As summarized in Table~\ref{tab:avsr}, we observe that the Whisper-based architecture (\textbf{S2}) benefits notably from explicit lip features, while LLM-based models (\textbf{S3}) show performance saturation when scaling individual components. Crucially, fusing the \textbf{S3-UASR} framework with the LLM-based model yields the best overall performance. 
This improvement highlights the complementarity between the \textbf{S3} and \textbf{S3-UASR} frameworks, outperforming homogeneous ensembling of \textbf{S3} models. 
Besides, posterior probability averaging consistently enhances robustness across all frameworks (\textbf{S1--S3}). 

\begin{table}
    \centering
    \vspace{-0.1cm}
    \caption{WER(\%) results on $\mathcal{D}'_{\mathrm{dev}}$. E1/E2 and E3/E4 were trained with different hyper-parameter settings.}
    \label{tab:avsr}
    \scalebox{0.78}{
    \begin{tabular}{c c c c  c }
    \hline
    Exp. ID  & Framework & Encoder  & Decoder &  WER (\%) \\
    \hline
    E1   & \multirow{4}{*}{AVSR S1} & \multirow{2}{*}{Face Large} &  
    \multirow{4}{*}{Transformer} & 36.71 \\
    E2    &  &  &  & 36.50 \\
    \cline{3-3}
    E3     &  & \multirow{2}{*}{Face\&Lip Large}  &  & 36.27 \\
    E4     &  &  &  & 36.51\\
    \hline
    --     & \multicolumn{3}{c}{Post. Prob. Avg. (E1+E2+E3+E4) }  & \textbf{34.92} \\
    \hline
    \multirow{2}{*}{E5}  & \multirow{4}{*}{AVSR S2} & Face Large + &  & \multirow{2}{*}{36.48} \\
       &                            & Whisper Encoder& Whisper  &   \\
       \cline{3-3}
    \multirow{2}{*}{E6}  &  & Face\&Lip Large + & Decoder  & \multirow{2}{*}{35.30} \\
       &                            & Whisper Encoder&  &   \\
    \hline
        --     & \multicolumn{3}{c}{Post. Prob. Avg. (E5+E6) }  & \textbf{34.21} \\
    \hline
    E7 & \multirow{3}{*}{AVSR S3} & Face Large & Qwen 2.5-7B & 35.64 \\
    \cline{3-4}
    E8 &  & Face Large & Qwen 2.5-14B & 35.72 \\
    \cline{3-4}
    E9 &  & Face\&Lip Large & Qwen 2.5-7B & 35.86 \\
    \hline
      --     & \multicolumn{3}{c}{Post. Prob. Avg. (E7+E8+E9) }  & \textbf{34.20} \\
    \hline
    \multirow{2}{*}{E10} & AVSR S3 & Face\&Lip Base +& \multirow{2}{*}{Qwen 2.5 7B} & \multirow{2}{*}{36.57} \\
        & -UASR   & WavLM          &              &       \\
    \hline
      --     & \multicolumn{3}{c}{Post. Prob. Avg. (E8+E10) }  & \textbf{33.62} \\
    \hline
    \end{tabular}
    }
\end{table}


\begin{table}[t]
\centering
\caption{Comparison of conversation clustering systems (Iterations $K$, Selection $M$, Clusters $N$).}
\label{tabscresult}
\renewcommand{\arraystretch}{0.7}
\begin{tabular}{l c c c c c} 
\toprule
Model & Para & $K$ & $M$ & $N$ & F1 Score (\%) \\
\midrule
Baseline & - & - & - & - & 81.53 \\
Qwen 2.5 & 72B & 10 & 10 & 5 & 98.89 \\
DeepSeek R1 & 671B & 10 & 10 & 5 & 100.00 \\
\midrule
Fusion & & & & & \textbf{100.00} \\
\bottomrule
\end{tabular}
\end{table}

For conversation clustering, the LLM-based conversation clustering approach based on \textbf{AVSR S1} results achieves high accuracy, and by fusing different large language models and the baseline method, the accuracy on the development set reaches 100$\%$ as presented in Table~\ref{tabscresult}.

\section{Conclusion}
The CHiME-9 MCoRec Challenge represents extensive speech overlap and concurrent multi-party dialogues, imposing substantial constraints for accurate ASR. We confronted these challenges via a integrated framework synergizing Audio-Visual Self-Supervised Learning (AV-SSL) and Large Language Models. Our pipeline cascades data-regularized AV-ASD offering precise temporal boundaries, semantic-aware AVTSE trained on extensive synthetic datasets, and a robust Whisper-LLM-based AVSR backend. Our key findings are: large-scale AV pre-training mitigates low SNR and overlap degradation; incorporating holistic facial features outperform lip-only representations for identity; pre-trained multi-modal representations substantially enhance detection, extraction, and ASR robustness, achieving 15.7\% JACER.

\section{Acknowledgements}
This work was supported by the National Natural Science Foundation of China under Grant No. U25A20409, National Natural Science Foundation of China under Grants No. 62401348, Fundamental Research Funds for the Central Universities under Grant No. GK202406005, and Young Talent Fund of Association for Science and Technology in Shaanxi, China (Grant No. 20250126).



\vfill\pagebreak


\begingroup
\small   
\bibliographystyle{IEEEbib}
\bibliography{strings,refs}

@article{nguyen2025cocktailtask,
  title={A Cocktail-Party Benchmark: Multi-Modal dataset and Comparative Evaluation Results},
  author={Nguyen, Thai-Binh and Zmolikova, Katerina and Ma, Pingchuan and Pham, Ngoc Quan and Fuegen, Christian and Waibel, Alexander},
  journal={arXiv preprint arXiv:2510.23276},
  year={2025}
}

@article{nguyen2025cocktailbenchmark,
  title={Cocktail-Party Audio-Visual Speech Recognition},
  author={Nguyen, Thai-Binh and Pham, Ngoc-Quan and Waibel, Alexander},
  journal={arXiv preprint arXiv:2506.02178},
  year={2025}
}

@article{lightASD,
       author = {{Liao}, Junhua and {Duan}, Haihan and {Feng}, Kanghui and {Zhao}, Wanbing and {Yang}, Yanbing and {Chen}, Liangyin},
        title = {A Light Weight Model for Active Speaker Detection},
      journal = {arXiv e-prints},
         year = 2023,
        month = mar,
          eid = {arXiv:2303.04439},
        pages = {arXiv:2303.04439},
          doi = {10.48550/arXiv.2303.04439},
archivePrefix = {arXiv},
       eprint = {2303.04439},
 primaryClass = {cs.CV},
       adsurl = {https://ui.adsabs.harvard.edu/abs/2023arXiv230304439L}
}

@article{AVA,
  author       = {Sourish Chaudhuri and
                  Joseph Roth and
                  Daniel P. W. Ellis and others},
  title        = {AVA-Speech: {A} Densely Labeled Dataset of Speech Activity in Movies},
  journal      = {CoRR},
  volume       = {abs/1808.00606},
  year         = {2018},
  url          = {http://arxiv.org/abs/1808.00606},
  eprinttype    = {arXiv},
  eprint       = {1808.00606},
  bibsource    = {dblp computer science bibliography, https://dblp.org}
}

@inproceedings{MSDWild,
  title     = {{MSDWild: Multi-modal Speaker Diarization Dataset in the Wild}},
  author    = {Tao Liu and Shuai Fan and Xu Xiang and others},
  year      = {2022},
  booktitle = {{Interspeech 2022}},
  pages     = {1476--1480},
  doi       = {10.21437/Interspeech.2022-10466},
  issn      = {2958-1796},
}

@article{M3SD,
  title={M3SD: Multi-modal, Multi-scenario and Multi-language Speaker Diarization Dataset},
  author={Wu, Shilong},
  journal={arXiv preprint arXiv:2506.14427},
  year={2025}
}

@inproceedings{dubey2023icassp,
  title={ICASSP 2023 Deep Noise Suppression Challenge},
  author={
 Dubey, Harishchandra and Aazami, Ashkan and Gopal, Vishak and Naderi, Babak and Braun, Sebastian and  Cutler, Ross and Gamper, Hannes and Golestaneh, Mehrsa and Aichner, Robert},
  booktitle={ICASSP},
  year={2023}
}

@inproceedings{haliassos2024braven,
  title={BRAVEn: Improving Self-supervised pre-training for Visual and Auditory Speech Recognition},
  author={Haliassos, Alexandros and Zinonos, Andreas and Mira, Rodrigo and Petridis, Stavros and Pantic, Maja},
  booktitle={ICASSP 2024-2024 IEEE International Conference on Acoustics, Speech and Signal Processing (ICASSP)},
  pages={11431--11435},
  year={2024},
  organization={IEEE}
}

@article{chen2021correlating,
  title={Correlating subword articulation with lip shapes for embedding aware audio-visual speech enhancement},
  author={Chen, Hang and Du, Jun and Hu, Yu and Dai, Li-Rong and Yin, Bao-Cai and Lee, Chin-Hui},
  journal={Neural Networks},
  volume={143},
  pages={171--182},
  year={2021},
  publisher={Elsevier}
}

@INPROCEEDINGS{659110,
  author={Fiscus, J.G.},
  booktitle={1997 IEEE Workshop on Automatic Speech Recognition and Understanding Proceedings}, 
  title={A post-processing system to yield reduced word error rates: Recognizer Output Voting Error Reduction (ROVER)}, 
  year={1997},
  volume={},
  number={},
  pages={347-354},
  doi={10.1109/ASRU.1997.659110}}

@inproceedings{radford2023robust,
  title={Robust speech recognition via large-scale weak supervision},
  author={Radford, Alec and Kim, Jong Wook and Xu, Tao and Brockman, Greg and McLeavey, Christine and Sutskever, Ilya},
  booktitle={International conference on machine learning},
  pages={28492--28518},
  year={2023},
  organization={PMLR}
}

@inproceedings{liu2022convnet,
  title={A convnet for the 2020s},
  author={Liu, Zhuang and Mao, Hanzi and Wu, Chao-Yuan and Feichtenhofer, Christoph and Darrell, Trevor and Xie, Saining},
  booktitle={Proceedings of the IEEE/CVF conference on computer vision and pattern recognition},
  pages={11976--11986},
  year={2022}
}

@inproceedings{gulati20_interspeech,
  title     = {{Conformer: Convolution-augmented Transformer for Speech Recognition}},
  author    = {Anmol Gulati and James Qin and Chung-Cheng Chiu and Niki Parmar and Yu Zhang and Jiahui Yu and Wei Han and Shibo Wang and Zhengdong Zhang and Yonghui Wu and Ruoming Pang},
  year      = {2020},
  booktitle = {{Interspeech 2020}},
  pages     = {5036--5040},
  doi       = {10.21437/Interspeech.2020-3015},
  issn      = {2958-1796},
}

@inproceedings{chiu2022self,
  title={Self-supervised learning with random-projection quantizer for speech recognition},
  author={Chiu, Chung-Cheng and Qin, James and Zhang, Yu and Yu, Jiahui and Wu, Yonghui},
  booktitle={International Conference on Machine Learning},
  pages={3915--3924},
  year={2022},
  organization={PMLR}
}

@inproceedings{zhang2022lip,
  title={Is lip region-of-interest sufficient for lipreading?},
  author={Zhang, Jing-Xuan and Wan, Genshun and Pan, Jia},
  booktitle={Proceedings of the 2022 International Conference on Multimodal Interaction},
  pages={368--372},
  year={2022}
}

@article{zhang2025target,
  title={Target speaker lipreading by audio--visual self-distillation pretraining and speaker adaptation},
  author={Zhang, Jing-Xuan and Mao, Tingzhi and Guo, Longjiang and Li, Jin and Zhang, Lichen},
  journal={Expert Systems with Applications},
  volume={272},
  pages={126741},
  year={2025},
  publisher={Elsevier}
}

@inproceedings{NEURIPS2022_960a172b,
 author = {Alayrac, Jean-Baptiste and Donahue, Jeff and Luc, Pauline and others},
 booktitle = {Advances in Neural Information Processing Systems},
 editor = {S. Koyejo and S. Mohamed and A. Agarwal and D. Belgrave and K. Cho and A. Oh},
 pages = {23716--23736},
 publisher = {Curran Associates, Inc.},
 title = {Flamingo: a Visual Language Model for Few-Shot Learning},
 url = {https://proceedings.neurips.cc/paper_files/paper/2022/file/960a172bc7fbf0177ccccbb411a7d800-Paper-Conference.pdf},
 volume = {35},
 year = {2022}
}

@article{zhang2025adapting,
  title={Adapting Speech Foundation Models with Large Language Models for Unified Speech Recognition},
  author={Zhang, Jing-Xuan and Wan, Genshun and Li, Jin and Gao, Jianqing},
  journal={arXiv preprint arXiv:2510.22961},
  year={2025}
}

@article{zhang2025audio,
  title={Audio-visual representation learning via knowledge distillation from speech foundation models},
  author={Zhang, Jing-Xuan and Wan, Genshun and Gao, Jianqing and Ling, Zhen-Hua},
  journal={Pattern Recognition},
  volume={162},
  pages={111432},
  year={2025},
  publisher={Elsevier}
}

@misc{qwen2025qwen25technicalreport,
      title={Qwen2.5 Technical Report}, 
      author={Team, Qwen and others},
      year={2025},
      eprint={2412.15115},
      archivePrefix={arXiv},
      primaryClass={cs.CL},
      url={https://arxiv.org/abs/2412.15115}, 
}

@article{hu2022lora,
  title={Lora: Low-rank adaptation of large language models.},
  author={Hu, Edward J and Shen, Yelong and Wallis, Phillip and Allen-Zhu, Zeyuan and Li, Yuanzhi and Wang, Shean and Wang, Lu and Chen, Weizhu and others},
  journal={ICLR},
  volume={1},
  number={2},
  pages={3},
  year={2022}
}

@article{afouras2018deep,
  title={Deep audio-visual speech recognition},
  author={Afouras, Triantafyllos and Chung, Joon Son and Senior, Andrew and Vinyals, Oriol and Zisserman, Andrew},
  journal={IEEE transactions on pattern analysis and machine intelligence},
  volume={44},
  number={12},
  pages={8717--8727},
  year={2018},
  publisher={IEEE}
}

@article{afouras2018lrs3,
  title={LRS3-TED: a large-scale dataset for visual speech recognition},
  author={Afouras, Triantafyllos and Chung, Joon Son and Zisserman, Andrew},
  journal={arXiv preprint arXiv:1809.00496},
  year={2018}
}

@InProceedings{Chung18b,
              author       = "Chung, J.~S. and Nagrani, A. and Zisserman, A.",
              title        = "VoxCeleb2: Deep Speaker Recognition",
              booktitle    = "INTERSPEECH",
              year         = "2018",
            }

@article{10.1145/3197517.3201357,
author = {Ephrat, Ariel and Mosseri, Inbar and Lang, Oran and Dekel, Tali and Wilson, Kevin and Hassidim, Avinatan and Freeman, William T. and Rubinstein, Michael},
title = {Looking to listen at the cocktail party: a speaker-independent audio-visual model for speech separation},
year = {2018},
issue_date = {August 2018},
publisher = {Association for Computing Machinery},
address = {New York, NY, USA},
volume = {37},
number = {4},
issn = {0730-0301},
url = {https://doi.org/10.1145/3197517.3201357},
doi = {10.1145/3197517.3201357},
journal = {ACM Trans. Graph.},
month = jul,
articleno = {112},
numpages = {11}
}

@inproceedings{nguyen25b_interspeech,
  title     = {{Cocktail-Party Audio-Visual Speech Recognition}},
  author    = {Thai-Binh Nguyen and Ngoc-Quan Pham and Alexander Waibel},
  year      = {2025},
  booktitle = {{Interspeech 2025}},
  pages     = {1828--1832},
  doi       = {10.21437/Interspeech.2025-676},
  issn      = {2958-1796},
}

@inproceedings{rouditchenko24_interspeech,
  title     = {{Whisper-Flamingo}: Integrating Visual Features into Whisper for Audio-Visual Speech Recognition and Translation},
  author    = {Andrew Rouditchenko and Yuan Gong and Samuel Thomas and Leonid Karlinsky and Hilde Kuehne and Rogerio Feris and James Glass},
  year      = {2024},
  booktitle = {Proceedings of the Annual Conference of the International Speech
                  Communication Association},
  pages     = {2420--2424},
}
\endgroup
\end{document}